\RequirePackage{lineno}
\documentclass[epj,nopacs]{svjour}
\usepackage{subfigure}
\usepackage{mathrsfs}
\usepackage{graphicx}
\usepackage{multirow}
\usepackage{rotating}
\usepackage{atlasphysics}
\usepackage{url}
\usepackage{hyperref}
\usepackage{textpos}
\usepackage{color}

\newcommand{\mct}{\ensuremath{m_\mathrm{CT}}}

\def\lsim{\mathrel{\rlap{\lower4pt\hbox{\hskip1pt$\sim$}}
    \raise1pt\hbox{$<$}}}                
\def\gsim{\mathrel{\rlap{\lower4pt\hbox{\hskip1pt$\sim$}}
    \raise1pt\hbox{$>$}}}                

\makeatletter
\renewcommand{\p@subfigure}{\arabic{figure}}
\renewcommand{\thesubfigure}{\alph{subfigure}}
\renewcommand{\@thesubfigure}{(\thesubfigure)}
\makeatother

\usepackage{setspace}
\titlerunning{Search for events with an identical flavour lepton pair and $\met$ at ATLAS}
\newcommand{\dileptitle}{Search for an excess of events with an identical flavour lepton pair and significant missing transverse momentum in $\sqrt{s}=7\TeV$ proton-proton collisions with the ATLAS detector}

\newcommand{\hideit}[1]{{ }}

\newcommand{\alpgen}{{\tt ALPGEN}}

\newcommand{\prospino}{{\tt PROSPINO}}
\newcommand{\herwig}{{\tt HERWIG}}
\newcommand{\pythia}{{\tt PYTHIA}}
\newcommand{\mcatnlo}{{\tt MC@NLO}}

\newcommand{\jimmy}{{\tt JIMMY}}

\newcommand{\ourintlumi}{{$35$~pb$^{-1}$}}

\begin{document}
\setlength\linenumbersep{2pt}

\title{\dileptitle \begin{textblock}{6}[0,1](0,-3){\small CERN-PH-EP-2011-039, Submitted to EPJC Letters}
\end{textblock}
}

\author{The ATLAS Collaboration}
\institute{}

\date{}

\abstract{
Results are presented of a search for supersymmetric particles decaying into final states with significant missing transverse momentum and exactly two identical flavour leptons ($e$, $\mu$) of opposite charge in $\sqrt{s}=7$~TeV collisions at the Large Hadron Collider. This channel is particularly sensitive to supersymmetric particle cascade decays producing flavour correlated lepton pairs. Flavour uncorrelated backgrounds are subtracted using a sample of opposite flavour lepton pair events. Observation of an excess beyond Standard Model expectations following this subtraction procedure would offer one of the best routes to measuring the masses of supersymmetric particles. In a data sample corresponding to an integrated luminosity of \ourintlumi ~no such excess is observed. Model-independent limits are set on the contribution to these final states from new physics and are used to exclude regions of a phenomenological supersymmetric parameter space.}

\maketitle

In this letter the first results are reported of a search for the production of supersymmetric (SUSY) \cite{susy} particles at ATLAS in events with exactly two leptons of identical flavour ($e$ or $\mu$) and opposite charge, and significant missing transverse momentum ($\MET$). This signature can be generated in SUSY events by the correlated production of leptons, for instance via the decay chains $\ninotwo \rightarrow \slepton \ell^{\pm} \rightarrow \ninoone \ell^{+} \ell^{-}$ or $\ninotwo \rightarrow \ninoone \ell^{+} \ell^{-}$.  Such events offer one of the best routes to model-independent measurements of the masses of SUSY particles via end-points in the lepton pair invariant mass distribution \cite{Hinchliffe:1996iu,Allanach:2000kt,Costanzo:2009mq} and so are of great interest especially if SUSY is found. The dominant sources of Standard Model (SM) background generally possess equal branching fractions for the production of lepton pairs of identical and different flavour, and can therefore be removed with a `flavour subtraction' procedure \cite{Hinchliffe:1996iu} in which the observation in the $e\mu$ channel is subtracted from that in the $ee$ and $\mu\mu$ channels. Specifically targeting this important technique for measuring SUSY particle masses, this analysis benefits from reduced sensitivity to systematic uncertainties in background estimates compared with other techniques. The results reported here are complementary to those of inclusive SUSY particle searches using lepton pairs \cite{atlas-2lepinc-pub}, and also to those of inclusive searches requiring jets, $\MET$ and zero leptons \cite{atlas-0lep-pub} or one lepton \cite{atlas-1lep-pub}. A search by CMS for SUSY in events with lepton pairs is reported in Ref.~\cite{Chatrchyan:2011bz}.

\vspace{-0.06cm}
The ATLAS detector~\cite{Aad:2008zzm} is a multipurpose particle
physics apparatus with a forward-backward symmetric cylindrical~geometry and near 4$\pi$ coverage in solid angle \footnote{ATLAS uses a right-handed coordinate system with its origin at the 
nominal interaction point (IP) in the centre of the detector 
and the $z$-axis coinciding with the axis of the beam pipe. 
The $x$-axis points from the IP to the centre of the LHC ring, 
and the $y$ axis points upward. Cylindrical coordinates 
$(r,\phi)$ are used in the transverse plane, $\phi$ being 
the azimuthal angle around the beam pipe. The pseudorapidity is 
defined in terms of the polar 
angle $\theta$ as $\eta=-\ln\tan(\theta/2)$.}.
The inner tracking detector (ID) consists of a
silicon pixel detector, a silicon microstrip detector (SCT), and
a transition radiation tracker (TRT) which also provides particle identification capability. 
The ID is surrounded by a thin superconducting
solenoid providing a 2 T magnetic field, and by
high-granularity liquid-argon (LAr) sampling electromagnetic calorimeters.
Hadronic coverage is provided by an iron-scintillator tile calorimeter in the central rapidity range.
The end-cap and forward regions 
are instrumented with LAr calorimetry
for both electromagnetic and hadronic measurements.
The muon spectrometer (MS) surrounds the calorimeters  and consists 
of three large superconducting toroids, a system 
of precision tracking chambers,
and detectors for triggering.

The $pp$-collision data used in this analysis were collected between March and November 2010 
at the LHC operating at a centre-of-mass energy of 7 \TeV. 
Application of basic beam, detector and data-quality requirements results in
a total integrated luminosity of \ourintlumi. The uncertainty on the
luminosity is estimated to be 11\%~\cite{atlas-lumi}.
The data have been collected with a single lepton ($e$ or $\mu$) trigger. The detailed
trigger requirements vary throughout the data-taking period due to the rapidly
increasing LHC luminosity and the commissioning of the trigger system, but always have a threshold that ensures a trigger efficiency for leptons
with transverse momentum $p_\mathrm{T} > 20$~GeV at the plateau. The efficiency of the triggers is studied with data,
and agrees well with expectations.

Monte Carlo (MC) simulated event samples are used to develop and validate the analysis procedure and to estimate the residual SM backgrounds following flavour subtraction. Samples of QCD jet events are generated with \pythia~\cite{Sjostrand:2006za}, using the {\tt MRST2007LO*} modified leading-order parton distribution functions (PDF)~\cite{Sherstnev:2007nd}, which are used with all leading-order (LO) MC codes.
Production of top quark pairs is simulated with \mcatnlo~\cite{Frixione:2002ik,Frixione:2003ei} (with a top quark mass of $172.5$ \GeV) and the next-to-leading order (NLO) PDF set {\tt CTEQ6.6}~\cite{Nadolsky:2008zw}, which is used with all NLO MC codes. Samples of $W$ and Z/$\gamma^*$ production with accompanying jets are produced
with \alpgen~\cite{Mangano:2002ea}. Diboson ($WW$, $WZ$, $ZZ$) production is simulated
with \herwig~\cite{Corcella:2000bw,herwig65long}, single top production 
with \mcatnlo~\cite{Frixione:2005vw,Frixione:2008yi}, and Drell-Yan with \pythia.
Fragmentation and hadronization for the \alpgen~and \mcatnlo~samples are performed with \herwig, using
\jimmy~\cite{Butterworth:1996zw} for the underlying event. The MC samples are produced using the ATLAS MC09 parameter 
tune \cite{mc09tunes} and a GEANT4~\cite{Agostinelli:2002hh} based detector simulation~\cite{:2010wqa}. 

Criteria for electron and muon identification closely follow those
described in Ref.~\cite{Aad:2010yt}.
Candidate electrons are required to pass ``tight'' electron selection
criteria and isolation requirements, and have $\pt > 20~\GeV \,$ and $|\eta| <~2.47$. Identified electrons are used to select events for both the signal region of the analysis and control regions used to estimate backgrounds. ``Medium'' electron selection criteria are mainly based on lateral shower shape requirements in the calorimeter, while $E/p$ (where $E$ is the shower energy in the calorimeter and $p$ the track momentum in the ID) and TRT cuts are applied for the tight electron selection, which provides additional rejection against conversions and fakes from hadrons. The electron isolation criteria require that the total transverse energy within a cone size $\Delta R=\sqrt{(\Delta\eta)^2+(\Delta\phi)^2}=0.2$ around the electron, is less than 0.15 of the electron $p_T$. Events are always vetoed if a medium electron is found in the transition region between the barrel and end-cap electromagnetic
calorimeter, $1.37 < |\eta| <~1.52$. 
Muons are required to be identified either in both the ID
and MS systems (combined muons) or as a match
between an extrapolated ID track and one or more track segments in the MS.
The ID track is required to have at least one pixel hit, more than five SCT hits, 
and a number of TRT hits that varies with $\eta$.
For combined muons, a good match between ID and MS tracks is required, and
the $\pt$ values measured by these two systems
must be compatible within the resolution. Isolation requirements are imposed, whereby the summed $\pt$ of other ID tracks above 500~MeV within a distance $\Delta R < 0.2$ 
around the muon track is required to be less than $1.8$ \GeV.
Only muons with $\pt > 20 \GeV$ and $|\eta| < 2.4$ are considered.
For the final selection, the distance between the $z$ coordinate of the 
primary vertex and that of the extrapolated muon track at the point
of closest approach to the primary vertex
must be less than $10$ mm. Jets are reconstructed using the anti-$k_t$ jet clustering 
algorithm~\cite{Cacciari:2008gp} with
a distance parameter $D = 0.4$. The inputs to this algorithm are clusters of
calorimeter cells seeded by cells with energy significantly above the measured
noise. Jets are constructed by performing a four-vector sum over
these clusters, treating each cluster as an $(E,\vec{p})$ four-vector with zero mass.
Jets are corrected for calorimeter non-compensation, material
and other effects using $\pt$- and $\eta$-dependent calibration
factors obtained from Monte Carlo and validated with test-beam and collision-data studies~\cite{atlascalibration}. 
Only jets with $\pt > 20~\GeV \,$
and $|\eta| < 2.5$ are considered. If a jet and a medium electron are both identified within a distance
$\Delta R < 0.2$ of each other, the jet is discarded.
Furthermore, identified medium electrons or muons are only considered if they 
satisfy $\Delta R > 0.4$ with respect to the closest remaining jet.
Events are discarded if they contain any jet failing
basic quality selection criteria, which rejects detector noise and 
non-collision backgrounds~\cite{atlas-jet-clean}. The calculation of the missing transverse momentum, $\met$, is based on the
modulus of the vector sum of the $\pt$ of the reconstructed objects (jets
with $\pt > 20 \GeV$, but over the full calorimeter coverage $|\eta| < 4.9$, 
and selected leptons), any additional non-isolated muons, and the
calorimeter clusters not belonging to reconstructed objects.

``Signal region'' events that contain lepton pairs of identical flavour ($e^+ e^-$ and $\mu^+ \mu^-$) and different flavour ($e^\pm \mu^\mp$) are selected, with the two populations subsequently used to calculate the excess of identical flavour events. Selected events must contain exactly two opposite sign leptons ($e$ or $\mu$), with invariant mass ($m_{\ell\ell}$) greater than 5~GeV. The $\MET$ must exceed 100~GeV in order to reject SM $Z+$jets events whilst maintaining efficiency for a range of SUSY models. Events must also possess at least one reconstructed primary vertex with at least five associated tracks. A flavour subtraction is performed through the use of the quantity $\mathcal{S}$ defined as
\small
\begin{equation}
\mathcal{S} = \frac{N(e^\pm e^\mp)}{\beta(1-(1-\tau_e)^2)} - \frac{N(e^\pm \mu^\mp)}{1-(1-\tau_e)(1-\tau_\mu)} + \frac{\beta N(\mu^\pm \mu^\mp)}{(1-(1-\tau_\mu)^2)},
\label{eq:FlavSubtraction}
\end{equation}
\normalsize
which measures the excess of identical-flavour events (first and third terms) over different-flavour events (second term), taking into account the electron and muon plateau trigger efficiencies ($\tau_e$ and $\tau_\mu$) and the ratio of electron to muon efficiency times acceptance ($\beta$). The trigger efficiencies for offline reconstructed objects are $\tau_e=$(98.5$\pm$1.1)$\%$ and $\tau_\mu=$(83.7$\pm$1.9)$\%$ respectively, while $\beta$ is determined from data to be 0.69$\pm$0.03, with the quoted errors including both systematic and statistical uncertainties. 

The value of $\mathcal{S}$ obtained from selected identical-flavour and different-flavour lepton SM events is expected to be small but non-zero, due primarily to $Z/\gamma^*$ boson production. The contributions to $\mathcal{S}$ expected from SM processes are estimated using a combination of Monte Carlo simulation and data-driven techniques. Contributions from single top and diboson events are estimated using the MC samples described above, scaled to the luminosity of the data sample. Contributions from $Z/\gamma^*+$jets, $t\bar{t}$ and events containing fake leptons (from QCD jets and $W+$jets events) are estimated using MC samples normalised to data in an appropriate control region. The $Z/\gamma^*$ control region contains lepton pair events satisfying the same selection criteria as the signal region but with $\MET<20$~GeV and an additional $81<m_{\ell\ell}<101$~GeV requirement. The $t\bar{t}$ control region~\cite{atlas-2lepinc-pub} contains ``top-tagged" lepton pair events again satisfying the same selection criteria as signal candidates but with $60<\MET<80$~GeV and an additional requirement of $\geq2$ jets with $p_\mathrm{T}>20$~GeV. The top-tagging requirement is imposed through the use of the variable $\mct$ \cite{Tovey:2008ui}, which can be calculated from the four-vectors of the selected jets and leptons:
\begin{equation}
  \mct^{2}(v_{1},v_{2}) = \left[ E_{\rm T}(v_{1}) + E_{\rm T}(v_{2}) \right]^2 -  \left[  {\bf p_{\rm T}}(v_{1}) - {\bf p_{\rm T}}(v_{2}) \right]^2 ,
  \label{eq:mct_endpoints}
\end{equation}
where $v_i$ can be a lepton, a jet, or a lepton-jet combination, transverse momentum vectors are denoted by ${\bf p_{\rm T}}$ and transverse energies $E_T$ are defined as $E_T = \sqrt{p_{\rm T}^2 +m^2}$. This quantity is bounded from above by analytical functions of the top quark and $W$ masses as described in Ref.~\cite{Polesello:2009rn}. Top-tagged events are required to possess $\mct$ values calculated from combinations of jets and leptons consistent with the expected bounds from $t\bar{t}$ events, as well as lepton-jet invariant mass values consistent with top quark decays. An electron control region for fake lepton events requires events to possess $\MET<60$~GeV, $\Delta\phi$ between the $\MET$ vector and a jet $<0.1$ and an electron with $p_\mathrm{T}>30$~GeV. A single muon control region for fake lepton events requires events to possess $\MET<30$~GeV, a muon with $p_\mathrm{T}<40$~GeV and a transverse mass $m_T(\mu,\MET)<30$~GeV. The electron and muon identification criteria are relaxed, to obtain a `looser' sample dominated by fakes. A loose-tight matrix method is then used to estimate the number of events with fake leptons in the signal region after final selection criteria. This method, which uses the probabilities derived from data for loosely selected
leptons and hadrons to satisfy the tight selection
criteria to predict the mixture of real and fake
leptons in the final sample, is similar to that described in Ref.~\cite{atlas-topxsec}.
The dominant uncertainties in the data-normalised background estimates arise from limited numbers of events in the control regions, theoretical uncertainties (including choice of generator, initial and final state radiation), an approximate $\sim\pm$7\% jet energy scale uncertainty~\cite{atlas-inclusivejet} and an approximate $\sim14\%$ jet energy resolution uncertainty~\cite{atlas-jer}. The latter uncertainties affect the shapes of the MC $\MET$ distributions. Uncertainties on backgrounds estimated solely with MC are dominated by the jet energy scale and resolution.
\begin{table}
\begin{center}
\caption{\label{table:signalregion} Expected numbers of SM background events in the signal region for each of the three lepton flavour combinations. The estimates are obtained using the procedures described in the text. The quoted error includes systematic and statistical uncertainties. The negative number of fakes predicted in the $e^\pm \mu^\mp$ channel is an artifact of the matrix method used to estimate this contribution. In the `Total SM', the number of fakes in this channel is taken to be zero.}
\begin{tabular}{c|c|c|c}
\hline
 & $e^\pm e^\mp$ & $e^\pm \mu^\mp$ & $\mu^\pm \mu^\mp$ \\
\hline
\hline
Data & 4 & 13 & 13 \\
\hline
$Z/\gamma^*+$jets&0.40$\pm$0.46&0.36$\pm$0.20&0.91$\pm$0.67\\ 
Dibosons&0.30$\pm$0.11&0.36$\pm$0.10&0.61$\pm$0.10\\
$t\bar{t}$&2.50$\pm$1.02&6.61$\pm$2.68&4.71$\pm$1.91\\ 
Single top&0.13$\pm$0.09&0.76$\pm$0.25&0.67$\pm$0.33\\
Fakes&0.31$\pm$0.21&-0.15$\pm$0.08&0.01$\pm$0.01\\
\hline
Total SM&3.64$\pm$1.24&8.08$\pm$2.78&6.91$\pm$2.20\\
\hline
 \end{tabular}
  \end{center}
\end{table}

\begin{figure}
\begin{center}
\includegraphics[width=0.50\textwidth]{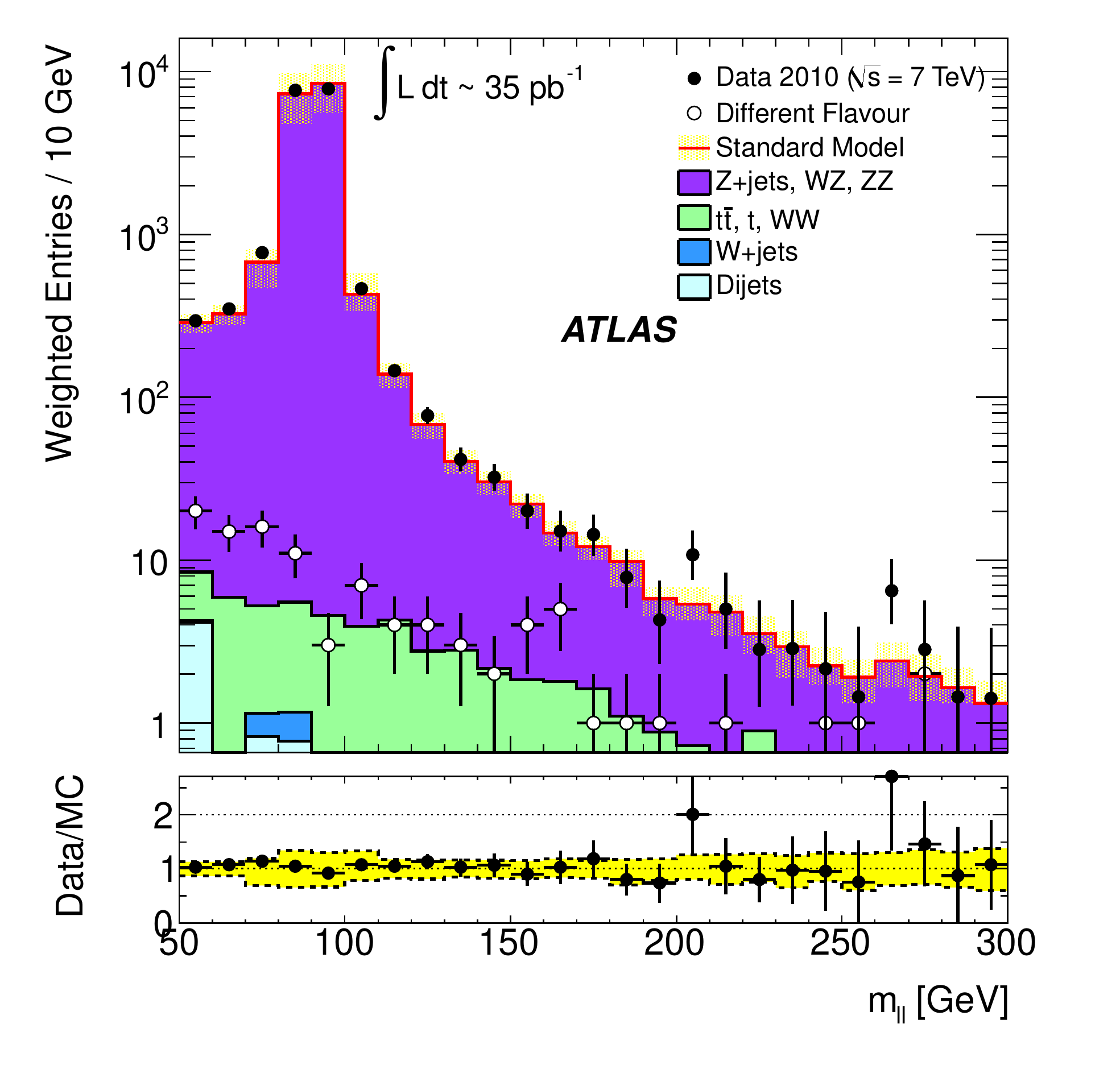}          
\caption{\label{fig:mll_main} Invariant mass distribution of identical flavour lepton pairs prior to applying the $\MET$ requirement, weighted by the acceptance and efficiency factors as in Equation~\ref{eq:FlavSubtraction}. The stacked histograms show the expected distributions from MC samples normalised to the luminosity of the data. The band indicates the uncertainty on the expectation from finite statistics, cross section, luminosity, jet and lepton energy scales and resolutions. Also shown is the observed distribution for different flavour pairs, weighted according to Equation~\ref{eq:FlavSubtraction}. In the region with $m_{ll}<100$~GeV, the dominant contributions to the different flavour data events are expected to come from $t\bar{t}$, QCD and $Z/\gamma^*+$jets events.}
\end{center}
\end{figure}

The invariant mass distributions of lepton pairs in selected data events, prior to applying the $\MET$ requirement, are presented in Figure~\ref{fig:mll_main}, weighted by the multiplicative factors in Equation~\ref{eq:FlavSubtraction} to yield the identical flavour and different flavour contributions to $\mathcal{S}$. After applying the $\MET>100$~GeV requirement  4, 13 and 13 events are observed in the $e^\pm e^\mp$, $e^\pm \mu^\mp$ and $\mu^\pm\mu^\mp$ channels respectively. The expected numbers of events in these channels, determined using the techniques described above, are listed in Table~\ref{table:signalregion} and are in reasonable agreement (within statistical and systematic uncertainties) with the observations. The dominant contribution to each channel arises from $t\bar{t}$ production. Using the observed numbers of events in each channel together with the measured values of $\tau_e$, $\tau_\mu$ and $\beta$, the observed value of $\mathcal{S}$ is found to be $\mathcal{S}_{obs}=1.98\pm0.15(\beta)\pm0.02(\tau_e)\pm0.06(\tau_\mu)$, where the uncertainties are those from the respective efficiency parameters. The expected mean value of $\mathcal{S}$ from SM background events alone is $\bar{\mathcal{S}_{b}}=2.06\pm0.79(stat.)\pm0.78(sys.)$, in good agreement with the observation. The dominant contributions to $\bar{\mathcal{S}_{b}}$ are from $Z/\gamma^*+$jets and diboson processes. The dominant contributions from $t\bar{t}$ events to the individual channels largely cancel when calculating $\mathcal{S}$, as expected. The $t\bar{t}$ population nevertheless has a significant impact on this analysis because the range of observed $\mathcal{S}$ values expected from a large number of hypothetical signal-free experiments is dominated by statistical fluctuations in the numbers of selected $t\bar{t}$ events in each channel.

To quantify the consistency between the oberved $\mathcal{S}$ value and the SM prediction the expected distribution of $\mathcal{S}_{b}$ in the absence of new physics must be determined. This distribution possesses a mean given by $\bar{\mathcal{S}_{b}}$ and a width dominated by statistical fluctuations in the numbers of events observed in each channel. The distribution can be determined by generating pseudo-experiments using the estimated mean numbers of background events from Table~\ref{table:signalregion} as input. For each pseudo-experiment the mean number of background events in each channel and from each source are sampled, taking appropriate account of correlations between the uncertainties in the estimates of these means. The resulting total mean number of background events in each channel is then used to construct a Poisson distribution from which the observed number of events in that channel is drawn. The resulting sampled event counts in each channel are then used with Equation~\ref{eq:FlavSubtraction}, taking care also to sample values of $\tau_e$, $\tau_\mu$ and $\beta$ according to their means and uncertainties, to determine a value of $\mathcal{S}_b$. The distribution of $\mathcal{S}_b$ values obtained in this way is used to estimate the probability of observing a value of $\mathcal{S}$ at least as large as $\mathcal{S}_{obs}$. 

\begin{figure}
\begin{center}
\includegraphics[width=0.5\textwidth]{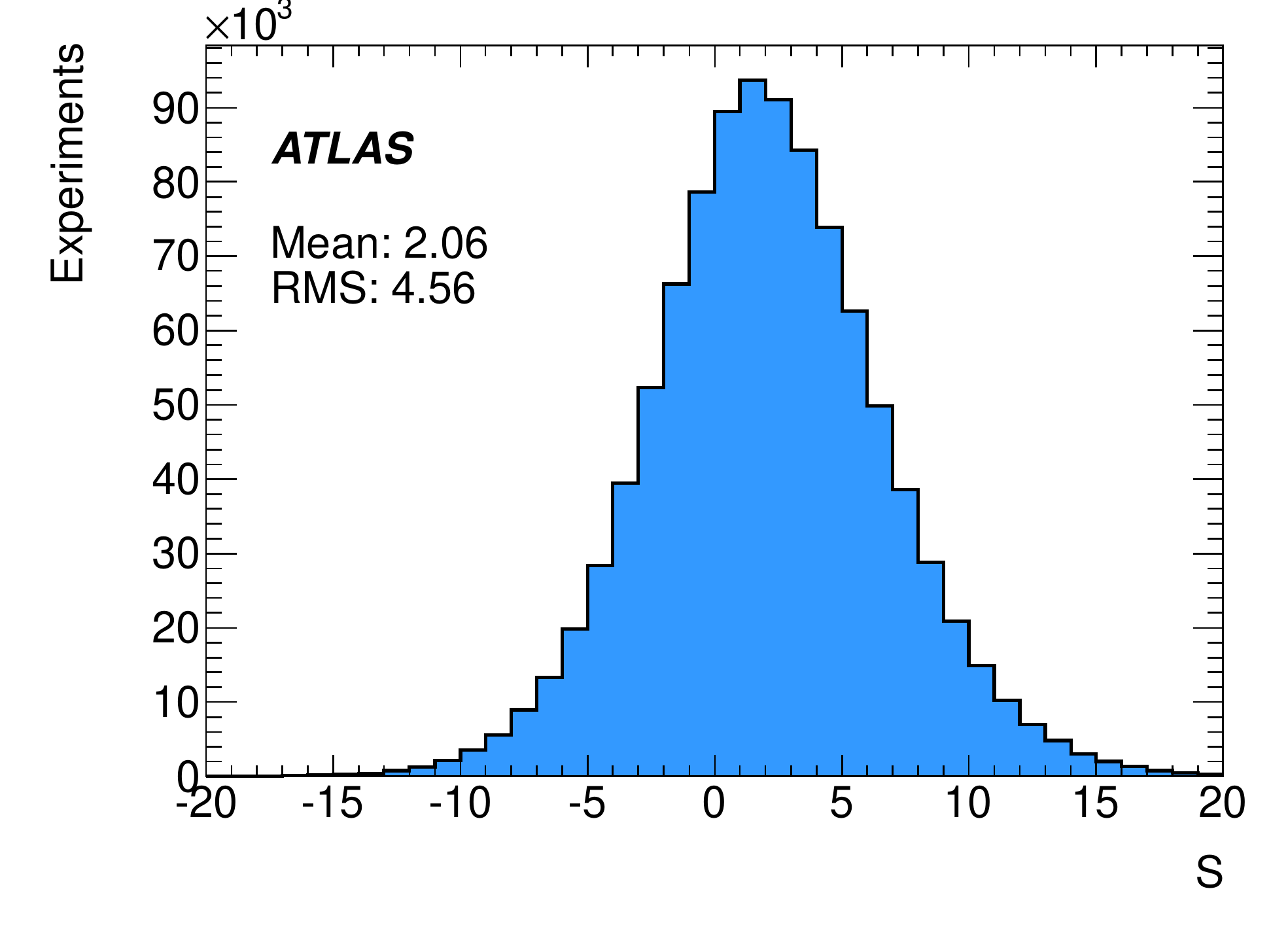}
\caption[Observed $\mathcal{S}$ for one-million hypothetical signal-free experiments]{\label{pValues:SMonly}Distribution of observed $\mathcal{S}$ values from one million hypothetical signal-free experiments. The shape is driven by statistical Poisson fluctuations in the expected rates of identical flavour and different flavour events, dominated by $t\bar{t}$ events.}
\end{center}
\end{figure}
The distribution of $\mathcal{S}$ values obtained from one million signal-free experiments using this procedure is shown in Figure~\ref{pValues:SMonly}. The shape of the distribution is dominated by statistical fluctuations in the numbers of events in each channel, with the uncertainty on $\bar{\mathcal{S}}_{b}$ being negligible by comparison. The probability of observing a value of $\mathcal{S}$ at least as large as $\mathcal{S}_{obs}$ is 49.7\% and hence no evidence of an excess of identical flavour events beyond SM expectations is observed.

Limits are set on $\bar{\mathcal{S}}_s$, the mean contribution to $\mathcal{S}$ from new physics. The statistical procedure employed follows that used to determine the consistency of the observed value of $\mathcal{S}$ with the background expectation. The pseudo-experiments are modified by adding signal event contributions to the input mean numbers of background events in each channel. An assumption must be made regarding the relative branching ratio of new physics events into identical flavour and different flavour channels, as adding flavour uncorrelated new physics contributions to the identical flavour and different flavour channels increases the width of the $\mathcal{S}$ distribution. Given such an assumption, a model-independent limit can be set on $\bar{\mathcal{S}}_s$ by comparing $\mathcal{S}_{obs}$ with the distribution of $\mathcal{S}$ values obtained from the new set of signal-plus-background pseudo-experiments.

If the assumption is made that the branching fractions for $e^\pm e^\mp$ and $\mu^\pm \mu^\mp$ final states in new physics events are identical, and the branching fraction for $e^\pm \mu^\mp$ final states is zero, a limit $\bar{\mathcal{S}_s}<8.8$ is set at 95$\%$ confidence level. Alternatively, if new physics events are assumed to possess a different flavour branching fraction of one half that for identical flavour events, then the limit becomes $\bar{\mathcal{S}_s}<12.6$ at 95$\%$ confidence. The limits are driven by the statistical fluctuations in $\mathcal{S}$, rather than systematic and statistical uncertainties in $\bar{\mathcal{S}_b}$ and in the variance of the $\mathcal{S}_b$ distribution. 

A similar procedure can be used to set limits within a specific new physics parameter space. In this case the mean numbers of signal events added to each channel are sampled according to the expectations from each point in the parameter space of the model together with the uncertainties in these expectations. The fraction of resulting pseudo-experiments with $\mathcal{S}<\mathcal{S}_{obs}$ gives the probability of the signal plus background hypothesis being falsely rejected. If the probability of being falsely rejected is $<5\%$, the point is excluded at 95$\%$ confidence.

\begin{figure}
\begin{center}
\includegraphics[width=0.50\textwidth]{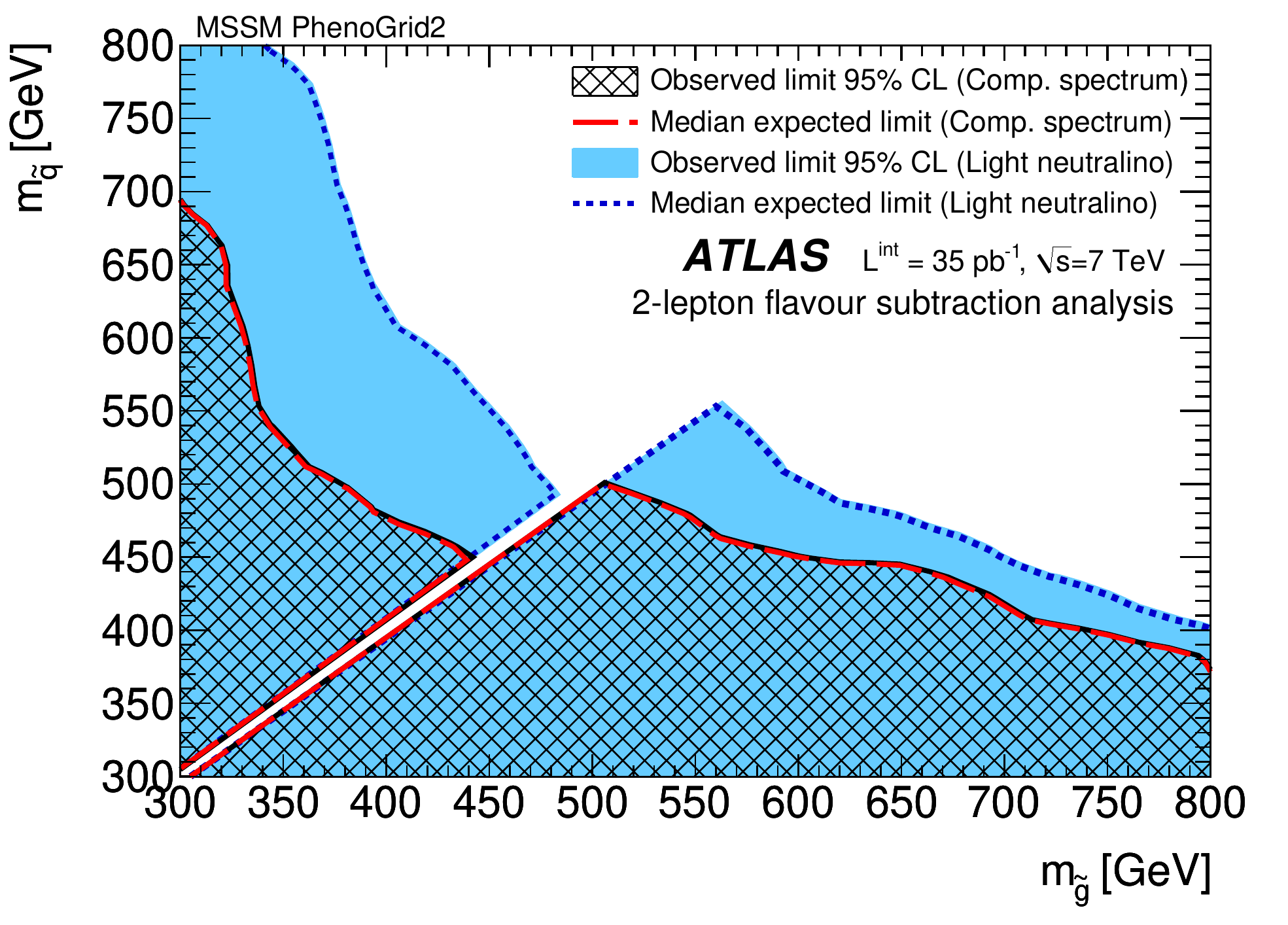}
\caption{\label{FinalFSLimitsSUGRA_main}Observed and expected 95\% C.L. exclusion limits in the $m_{\tilde{g}}-m_{\tilde{q}}$ plane for 24-parameter MSSM models. }
\end{center}
\end{figure}

As an example, two-dimensional grids in the parameter space of a 24 parameter MSSM model~\cite{Paige:2003mg} are considered (to be referred to as `MSSM PhenoGrid2'). The 24 parameter MSSM is a generic MSSM on which flavour and CP violation have been imposed. For these grids the following parameters are fixed: $m_A=1000$ GeV, $\mu = 1.5\min(m_{\tilde{g}},m_{\tilde{q}})$,  $\tan{\beta}=4$,  
$A_t$=$\mu/\tan{\beta}$, $A_b=\mu\tan{\beta}$, and $A_l=\mu\tan{\beta}$.
The masses of the $3^{rd}$ generation sfermions are set to 2~TeV, and common squark mass and slepton mass parameters are assumed for the first 
two generations.
Two grids in the $m_{\tilde{g}}-m_{\tilde{q}}$ plane are studied:
one with  a compressed spectrum yielding a soft final state kinematics, defined by
$m_{\tilde{\chi}_2^0}=M-50$ GeV, $m_{\tilde{\chi}_1^0}=M-150$ GeV and $m_{\tilde{l}_L}=M-100$ GeV, where $M$ is the minimum of the gluino and squark mass (`compressed spectrum'); and one with a very light LSP, yielding a harder spectrum of leptons, jets and $\met$,  with $m_{\tilde{\chi}_2^0}=M-100$ GeV,
$m_{\tilde{\chi}_1^0}=100$ GeV and $m_{\tilde{l}_L}=M/2$ GeV (`light neutralino'). Signal events are generated with \herwig ~for the MSSM grids. The cross sections are calculated
at NLO with \prospino~\cite{Beenakker:1996ch}. Theoretical and experimental uncertainties are determined for each model and used when sampling the mean numbers of signal events in each channel. Theoretical 
uncertainties are evaluated by varying the factorisation and renormalisation
scales and the {\tt CTEQ6.6} PDF sets~\cite{Nadolsky:2008zw} used for the cross section
calculation. Experimental uncertainties are dominated by the uncertainty on the jet energy scale and resolution. An 11\% luminosity uncertainty is included.
The results are shown in the $m_{\tilde{g}}-m_{\tilde{q}}$ plane in Figure~\ref{FinalFSLimitsSUGRA_main}. 
For `compressed spectrum' (`light neutralino') models and $m_{\tilde{g}}=m_{\tilde{q}}+10$~GeV, the 95\% confidence lower limit on $m_{\tilde{q}}$ is 503 (558)~GeV.

In summary, a flavour subtraction technique has been used to search for an excess beyond SM expectations of high missing transverse momentum events containing opposite charge identical flavour lepton pairs. No significant excess has been observed, allowing limits to be set on the model-independent quantity $\bar{\mathcal{S}}_s$, which measures the mean excess from new physics taking into account flavour-dependent acceptances and efficiencies.

We thank CERN for the very successful operation of the LHC, as well as the
support staff from our institutions without whom ATLAS could not be
operated efficiently.

We acknowledge the support of ANPCyT, Argentina; YerPhI, Armenia; ARC,
Australia; BMWF, Austria; ANAS, Azerbaijan; SSTC, Belarus; CNPq and FAPESP,
Brazil; NSERC, NRC and CFI, Canada; CERN; CONICYT, Chile; CAS, MOST and
NSFC, China; COLCIENCIAS, Colombia; MSMT CR, MPO CR and VSC CR, Czech
Republic; DNRF, DNSRC and Lundbeck Foundation, Denmark; ARTEMIS, European
Union; IN2P3-CNRS, CEA-DSM/\-IRFU, France; GNAS, Georgia; BMBF, DFG, HGF, MPG
and AvH Foundation, Germany; GSRT, Greece; ISF, MINERVA, GIF, DIP and
Benoziyo Center, Israel; INFN, Italy; MEXT and JSPS, Japan; CNRST, Morocco;
FOM and NWO, Netherlands; RCN, Norway; MNiSW, Poland; GRI\-CES and FCT,
Portugal; MERYS (MECTS), Romania; MES of Russia and ROSATOM, Russian
Federation; JINR; MSTD, Serbia; MSSR, Slovakia; ARRS and MVZT, Slovenia;
DST/\-NRF, South Africa; MICINN, Spain; SRC and Wallenberg Foundation,
Sweden; SER, SNSF and Cantons of Bern and Geneva, Switzerland; NSC, Taiwan;
TAEK, Turkey; STFC, the Royal Society and Leverhulme Trust, United Kingdom;
DOE and NSF, United States of America.

The crucial computing support from all WLCG partners is acknowledged
gratefully, in particular from CERN and the ATLAS Tier-1 facilities at
TRIUMF (Canada), NDGF (Denmark, Norway, Sweden), CC-IN2P3 (France),
KIT/GridKA (Germany), INFN-CNAF (Italy), NL-T1 \-(Netherlands), PIC (Spain),
ASGC (Taiwan), RAL (UK) and BNL (USA) and in the Tier-2 facilities
worldwide.

\bibliographystyle{atlasnote}
\bibliography{PUB_2lepton2010_fs}

\clearpage
\newpage
\begin{flushleft}
{\Large The ATLAS Collaboration}

\bigskip

G.~Aad$^{\rm 48}$,
B.~Abbott$^{\rm 111}$,
J.~Abdallah$^{\rm 11}$,
A.A.~Abdelalim$^{\rm 49}$,
A.~Abdesselam$^{\rm 118}$,
O.~Abdinov$^{\rm 10}$,
B.~Abi$^{\rm 112}$,
M.~Abolins$^{\rm 88}$,
H.~Abramowicz$^{\rm 153}$,
H.~Abreu$^{\rm 115}$
% ...
\bigskip

% ...
$^{10}$ Institute of Physics, Azerbaijan Academy of Sciences, Baku, Azerbaijan\\
$^{11}$ Institut de F\'isica d'Altes Energies and Universitat Aut\`onoma  de Barcelona and ICREA, Barcelona, Spain\\
% ...
$^{48}$ Fakult\"{a}t f\"{u}r Mathematik und Physik, Albert-Ludwigs-Universit\"{a}t, Freiburg i.Br., Germany\\
$^{49}$ Section de Physique, Universit\'e de Gen\`eve, Geneva, Switzerland\\
% ...
$^{88}$ Department of Physics and Astronomy, Michigan State University, East Lansing MI, United States of America\\
% ...
$^{111}$ Homer L. Dodge Department of Physics and Astronomy, University of Oklahoma, Norman OK, United States of America\\
$^{112}$ Department of Physics, Oklahoma State University, Stillwater OK, United States of America\\
% ...
$^{115}$ LAL, Univ. Paris-Sud and CNRS/IN2P3, Orsay, France\\
% ...
$^{118}$ Department of Physics, Oxford University, Oxford, United Kingdom\\
% ...
$^{153}$ Raymond and Beverly Sackler School of Physics and Astronomy, Tel Aviv University, Tel Aviv, Israel\\
% ...
\end{flushleft}

\end{document}